\documentclass[prd,superscriptaddress,showpacs,nofootinbib,amsmath,amssymb,aps,11pt]{revtex4}

\usepackage{bm}
\usepackage{amsfonts}
\usepackage{latexsym}
\usepackage[latin1]{inputenc}
\usepackage{graphicx}
\usepackage{amsmath}
\usepackage{palatino}
\usepackage{mathpazo}
\usepackage{booktabs}
\usepackage{dcolumn}

\def\jnl@style{\it}
\def\aaref@jnl#1{{\jnl@style#1}}

\def\aaref@jnl#1{{\jnl@style#1}}

\def\aj{\aaref@jnl{AJ}}                   
\def\apj{\aaref@jnl{ApJ}}                 
\def\apjl{\aaref@jnl{ApJ}}                
\def\apjs{\aaref@jnl{ApJS}}               
\def\apss{\aaref@jnl{Ap\&SS}}             
\def\aap{\aaref@jnl{A\&A}}                
\def\aapr{\aaref@jnl{A\&A~Rev.}}          
\def\aaps{\aaref@jnl{A\&AS}}              
\def\mnras{\aaref@jnl{Mon.~Not.~Roy.~Astron.~Soc.}}             
\def\prd{\aaref@jnl{Phys.~Rev.~D}}        
\def\prc{\aaref@jnl{Phys.~Rev.~C}}  
\def\prl{\aaref@jnl{Phys.~Rev.~Lett.}}    
\def\qjras{\aaref@jnl{QJRAS}}             
\def\skytel{\aaref@jnl{S\&T}}             
\def\ssr{\aaref@jnl{Space~Sci.~Rev.}}     
\def\zap{\aaref@jnl{ZAp}}                 
\def\nat{\aaref@jnl{Nature}}              
\def\aplett{\aaref@jnl{Astrophys.~Lett.}} 
\def\apspr{\aaref@jnl{Astrophys.~Space~Phys.~Res.}} 
\def\physrep{\aaref@jnl{Phys.~Rep.}}      
\def\physscr{\aaref@jnl{Phys.~Scr}}       
\def\commat{\aaref@jnl{Comm.~Math.~Phys.}}              
\def\science{\aaref@jnl{Science}}               
\def\cqg{\aaref@jnl{Classical Quant.~Grav.}}            
\def\jpcs{\aaref@jnl{JPCS}}                                     
\def\ijmpd{\aaref@jnl{Int.~J.~Mod.~Phys.~D}}                    
\def\grg{\aaref@jnl{Gen.~Relat.~Gravit.}}               
\def\rpp{\aaref@jnl{Rep.~Prog.~Phys.}}          
\def\npa{\aaref@jnl{Nucl.~Phys.~A}}        
\def\lrr{\aaref@jnl{Living Rev.~Rel.}}                   
\def\jcap{\aaref@jnl{J.~Cosmology Astropart.~Phys.}}    
\def\rmp{\aaref@jnl{Rev.~Mod.~Phys.}}   

\allowdisplaybreaks[1]

\begin{document}

\title{Slowly rotating neutron stars in scalar-tensor theories with a massive scalar field}

\author{Stoytcho S. Yazadjiev}
\email{yazad@phys.uni-sofia.bg}
\affiliation{Department of Theoretical Physics, Faculty of Physics, Sofia University, Sofia 1164, Bulgaria}

\author{Daniela D. Doneva}
\email{daniela.doneva@uni-tuebingen.de}
\affiliation{Theoretical Astrophysics, Eberhard Karls University of T\"ubingen, T\"ubingen 72076, Germany}
\affiliation{INRNE - Bulgarian Academy of Sciences, 1784  Sofia, Bulgaria}

\author{Dimitar Popchev}
\affiliation{Department of Theoretical Physics, Faculty of Physics, Sofia University, Sofia 1164, Bulgaria}


\begin{abstract}
In the scalar-tensor theories with a massive scalar field the coupling constants, and the coupling functions in general, which  are observationally allowed, can differ significantly from those in the massless case. This fact naturally implies that the scalar-tensor neutron stars with a massive scalar field can have rather different structure and properties in comparison with
their counterparts in the massless case and in general relativity. In the present paper we study slowly rotating neutron stars in scalar-tensor
theories with a massive gravitational scalar. Two examples of scalar-tensor theories are examined - the first example is the massive Brans-Dicke
theory  and the second one is a massive scalar-tensor theory indistinguishable from general relativity in the weak field limit. In the later case
we study the effect of the scalar field mass on the spontaneous scalarization of neutron stars. Our numerical results show  that the inclusion of
a mass term for the scalar field indeed changes the picture drastically compared to the massless case. It turns out
that mass, radius and moment of inertia for neutron stars in massive scalar-tensor theories can differ drastically from the pure general relativistic solutions if sufficiently large masses of the scalar field are considered.
\end{abstract}


\maketitle

\section{Introduction}

Scalar-tensor theories of gravity (STT) are the most natural viable generalization of general relativity (GR)
and have been extensively studied in various astrophysical and cosmological aspects in the last two decades.
Ones of the best laboratories for testing the strong field regime of scalar-tensor theories, and the gravitational theories in general,  are the neutron stars.
The scalar-tensor neutron stars attracted a lot of interest in the past and their structure, properties and the physical effects related to them were investigated in
many papers (\cite{Damour1993,Damour1996,Harada1997,Harada1998,Salgado1998,Pani2011,Sotani2012,Doneva2013} and references therein). Most of these studied were restricted to scalar-tensor theories with a massless scalar field. The recent astrophysical and cosmological observations, however,
have severely constrained the basic parameters of the scalar-tensor theories with a massless scalar field \cite{Freire2012,Antoniadis13} leaving a narrow window for new  physics beyond general relativity.
The situation can change drastically if we consider a massive scalar field. The scalar field mass $m_{\varphi}$ leads to a finite range of the scalar field of the order of its Compton wave-length $\lambda_\varphi=2\pi/m_{\varphi}$. In other words the presence of the scalar field will be suppressed outside the compact objects at distances $D>\lambda_{\varphi}$.
This means in turn that all observations of compact objects  involving distances greater than $\lambda_{\varphi}$ can not put constraints, or at least stringent constraints, on the scalar tensor theories. For example, in the case of massive Brans-Dicke theory with $m_{\varphi}\gtrsim 2\times 10^{-25} GeV$ (or $\lambda_{\varphi}\lesssim 10^{11} m $), the Solar System observations can not put stringent constraints on the Brans-Dicke parameter $\omega_{BD}$  and all values $\omega_{BD}>-\frac{3}{2}$ are observationally allowed \cite{Perivolaropoulos2010}. The massive gravitational scalar suppresses also the dipole radiation and the compact binaries can not constrain severely the Brans-Dicke parameter if their orbit radius is significantly greater
than $\lambda_{\varphi}$ \cite{Alsing2012}. In general, as shown in \cite{Alsing2012},  if $\lambda_{\varphi} << 10^{11} m$ ($\lambda_{\varphi} << 1 AU$) then $\omega_{BD}$ can take on any value as long as
$\omega_{BD}>-\frac{3}{2}$.

It is well known that certain scalar-tensor theories with a coupling function in the Einstein frame of the type $\alpha(\varphi)=\beta\varphi$ with $\beta<0$ exhibit the
non-perturbative effect of spontaneous scalarization for neutron stars  which consists in the fact that the scalar vacuum is unstable to the condensation of the scalar field in the present of matter. The present observations, for example  the pulsar-white dwarf binary PSR J0348+0432,  put very stringent bound on the coupling parameter $\beta$, namely $\beta \gtrsim -4.5$. However, if we consider a massive scalar field with a suitable range then the values of $\beta$ allowed by the observations can be in orders of magnitude different  from  $-4.5$. A rough estimate for the parameter $\beta$ when the range of the scalar field is significantly smaller than the periapse of  PSR J0348+0432 ($\lambda_{\varphi} << 10^{10} m$ or equivalently $m_{\varphi}>>10^{-16} eV$ ) was given in \cite{Ramazanouglu2016}, namely $3\lesssim-\beta \lesssim 10^3$.

As we saw above the coupling constants (and the coupling functions in general) of the scalar-tensor theories with a massive scalar field, which are observationally
allowed, can differ really significantly from those in the massless case.  This fact naturally leads us to the conclusion that the compact objects in general, and the neutron stars
in particular,  with a massive scalar field could  in principle have rather different structure and properties in comparison with their counterparts in the massless case.
With this motivation in mind, in the present paper we numerically study slowly rotating  neutron stars in scalar tensor theories with a massive scalar field. More precisely we
consider slowly rotating neutron star models in the massive Brans-Dicke theory and in the massive scalar-tensor theory given by  the Einstein frame coupling function
\begin{equation} \label{eq:coupling_function}
\alpha(\varphi)=\beta\varphi
\end{equation}
with $\beta<0$. In the later case we  study the spontaneous scalarization for massive scalar fields. To the best of our knowledge
the spontaneous scalarization with a massive scalar field for static and slowly rotating neutron stars was studied in the master thesis of one of us (D.P.) \cite{Popchev2015}.
During the preparation of the present work the nice paper \cite{Ramazanouglu2016} appeared where the authors also study the  spontaneous scalarization with a massive scalar field for
static (nonrotating) neutron stars. Let us note however that the mass-radius and  mass-moment of inertia relations are discussed only in  \cite{Popchev2015} and in the present paper. 
A model similar to the spontaneous scalarization (the so-called asymmetron model) was also considered in \cite{Chen2015}. The differences between the  scalar-tensor spontaneous scalarization 
and the spontaneous scalarization in the mentioned model were discussed in \cite{Ramazanouglu2016}.

\section{Basic equations}
The Einstein frame action of the scalar-tensor theories is given by

\begin{eqnarray}\label{EFA}
S=\frac{1}{16\pi G} \int d^4x \sqrt{-g}\left[ R - 2
g^{\mu\nu}\partial_{\mu}\varphi \partial_{\nu}\varphi - V(\varphi)
\right] + S_{\rm
matter}(A^2(\varphi)g_{\mu\nu},\chi),
\end{eqnarray}
where $R$ is the Ricci scalar curvature with respect to the
Einstein frame metric $g_{\mu\nu}$. The scalar-tensor theories are fully specified by the functions $A(\varphi)$
and $V(\varphi)$. The Jordan frame metric ${\tilde g}_{\mu\nu}$ and the gravitational scalar $\Phi$ are  given respectively by
${\tilde g}_{\mu\nu}= A^2(\varphi)g_{\mu\nu}$ and $\Phi=A^{-2}(\varphi)$. In the present paper we shall restrict ourselves to the following simple dilaton potential
$V(\varphi)=2m^2_{\varphi}\varphi^2$  yielding the mass of $\varphi$.

The field equations that follow from the action (\ref{EFA}) are

\begin{eqnarray}
&&R_{\mu\nu} - \frac{1}{2}g_{\mu\nu}R= 8\pi G T_{\mu\nu} + 2\nabla_{\mu}\varphi\nabla_{\nu}\varphi - g_{\mu\nu} g^{\alpha\beta}\nabla_{\alpha}\varphi\nabla_{\beta}\varphi
- \frac{1}{2}V(\varphi) g_{\mu\nu}, \\
&&\nabla_{\mu}\nabla^{\mu}\varphi = -4\pi G \alpha(\varphi) T + \frac{1}{4}\frac{dV(\varphi)}{d\varphi},
\end{eqnarray}
where $\nabla_{\mu}$  is the covariant derivative with respect to $g_{\mu\nu}$  and the coupling function $\alpha(\varphi)$ is defined by $\alpha(\varphi)=\frac{d\ln A(\varphi)}{d\varphi}$. From the field equations and the contracted Bianchi identities
we find the following conservation law for the Einstein frame energy-momentum tensor

\begin{eqnarray}
\nabla_{\mu}T^{\mu}{\nu}= \alpha(\varphi) T\nabla_{\nu}\varphi .
\end{eqnarray}

The Einstein frame energy-momentum tensor $T_{\mu\nu}$  and the Jordan frame one ${\tilde T}_{\mu\nu}$ are related via the formula $T_{\mu\nu}=A^2(\varphi){\tilde T}_{\mu\nu}$.
In the case of a perfect fluid the relations between the energy density, pressure and 4-velocity in both frames are given by $\rho=A^4(\varphi){\tilde \rho}$, $p=A^4(\varphi){\tilde p}$
and $u_{\mu}=A^{-1}(\varphi){\tilde u}_{\mu}$.

As we discussed in the introduction we will concentrate on two classes of STT. The first one is the massive Brans-Dicke theory with a coupling function
\begin{equation}\label{eq:BD_couplingFunction}
\alpha(\varphi)=\alpha_0  \Leftrightarrow   A(\varphi)=\exp(\alpha_{0}\varphi),
\end{equation}
where $\alpha_0$ is a constant. The second one is a massive scalar-tensor theory with
\begin{equation}
\alpha(\varphi)=\beta\varphi   \Leftrightarrow A(\varphi)=\exp(1/2\varphi^2),
\end{equation}
where $\beta<0$ is a parameter. The latter case is equivalent to general relativity in the weak field regime.

We consider further stationary and axisymmetric spacetimes as well as stationary and axisymmetric  fluid and scalar field configurations.
In the slowly rotating approximation, i.e. keeping only first-order terms in the angular velocity $\Omega=u^{\phi}/u^{t}$, the spacetime metric
can be written in the standard form \cite{Hartle1967}

\begin{eqnarray}
ds^2= - e^{2\phi(r)}dt^2 + e^{2\Lambda(r)}dr^2 + r^2(d\theta^2 +
\sin^2\theta d\vartheta^2 ) - 2\omega(r,\theta)r^2 sin^2\theta  d\vartheta dt.
\end{eqnarray}

Only the metric function $\omega$ is in linear order of $\Omega$. The influence of the rotation on the other metric functions, the scalar field,
the fluid energy density and pressure is of order ${\cal O}(\Omega^2)$. For the fluid four-velocity  $u^{\mu}$, up to linear terms in $\Omega$, one finds
$u=u^{t}(1, 0, 0, \Omega)$, where $u^{t}=e^{-\phi(r)}$.

The dimensionally reduced Einstein frame field equations  containing at most terms linear in $\Omega$,  are the following

\begin{eqnarray}
&&\frac{1}{r^2}\frac{d}{dr}\left[r(1- e^{-2\Lambda})\right]= 8\pi G
A^4(\varphi) {\tilde \rho} + e^{-2\Lambda}\left(\frac{d\varphi}{dr}\right)^2
+ \frac{1}{2} V(\varphi), \label{eq:FieldEq1} \\
&&\frac{2}{r}e^{-2\Lambda} \frac{d\phi}{dr} - \frac{1}{r^2}(1-
e^{-2\Lambda})= 8\pi G A^4(\varphi) {\tilde p} +
e^{-2\Lambda}\left(\frac{d\varphi}{dr}\right)^2 - \frac{1}{2}
V(\varphi),\label{eq:FieldEq2}\\
&&\frac{d^2\varphi}{dr^2} + \left(\frac{d\phi}{dr} -
\frac{d\Lambda}{dr} + \frac{2}{r} \right)\frac{d\varphi}{dr}= 4\pi G
\alpha(\varphi)A^4(\varphi)({\tilde \rho}-3{\tilde p})e^{2\Lambda} + \frac{1}{4}
\frac{dV(\varphi)}{d\varphi} e^{2\Lambda}, \label{eq:FieldEq3}\\
&&\frac{d{\tilde p}}{dr}= - ({\tilde \rho} + {\tilde p}) \left(\frac{d\phi}{dr} +
\alpha(\varphi)\frac{d\varphi}{dr} \right), \label{eq:FieldEq4} \\
&&\frac{e^{\Phi-\Lambda}}{r^4} \partial_{r}\left[e^{-(\Phi + \Lambda)} r^4 \partial_{r}{\bar\omega} \right]  + \frac{1}{r^2\sin^3\theta} \partial_{\theta}\left[\sin^3\theta\partial_{\theta}\bar\omega \right]= 16\pi GA^4(\varphi)({\tilde \rho} + {\tilde p})\bar\omega ,
\end{eqnarray}
where the function $\bar\omega$ is defined as $\bar\omega = \Omega - \omega$.

This system of equations,  supplemented with the equation of state for the star matter and appropriate boundary conditions, describes the interior  and the exterior of
the neutron star. In the exterior of the neutron star we have to set ${\tilde \rho}={\tilde p}=0$.

The equation for $\bar\omega$ is separated from the other equations which form an independent subsystem.
This subsystem is obviously  the system for the static and spherically symmetric case. The natural boundary conditions
at the center of the star are $\rho(0)=\rho_{c}, \Lambda(0)=0,$  while at infinity we have $\lim_{r\to \infty}\phi(r)=0, \lim_{r\to \infty}\varphi
(r)=0$ as required by the asymptotic flatness \cite{Yazadjiev2014}. As usual,  the coordinate radius $r_S$ of the star is determined by the
condition $p(r_S)=0$ while  the physical radius of the star as measured in the physical Jordan frame is given by $R_{S}= A[\varphi(r_S)] r_S$.

The equation for $\bar \omega$ can be considerably simplified. Expanding $\bar \omega$ in the form \cite{Hartle1967}

\begin{eqnarray}
\bar\omega= \sum^{\infty}_{l=1}{\bar \omega}_{l}(r) \left(- \frac{1}{\sin\theta}\frac{dP_{l}}{d\theta} \right),
\end{eqnarray}
where $P_{l}$ are Legendre polynomials and substituting into the equation for $\bar \omega$ we find

\begin{eqnarray}\label{OL}
\frac{e^{\Phi-\Lambda}}{r^4} \frac{d}{dr}\left[e^{-(\Phi+ \Lambda)}r^4 \frac{d{\bar\omega}_{l}(r)}{dr} \right] - \frac{l(l+1)-2}{r^2} {\bar\omega}_{l}(r)=
16\pi G A^4(\varphi)(\rho + p){\bar\omega}_{l}(r).
\end{eqnarray}

 For asymptotically flat spacetimes, the asymptotic of the exterior solution of  (\ref{OL}) is ${\bar \omega}_{l} \to {\rm const}_1\, r^{-l-2} + {\rm const}_2\, r^{l-1}$. In view of the fact that $\omega \to 2J/r^3$ (or equivalently $\bar\omega \to \Omega - 2J/r^3 $) for $r\to \infty$ with $J$ being the angular momentum of the star and comparing it with the above asymptotic for $\bar \omega$, we conclude that $l=1$, i.e. ${\bar \omega}_{l}=0$ for $l\ge 2$. Therefore $\bar\omega$ is a function of $r$ only and the equation for $\bar \omega$  is

\begin{eqnarray}\label{OR}
\frac{e^{\Phi-\Lambda}}{r^4} \frac{d}{dr}\left[e^{-(\Phi+ \Lambda)}r^4 \frac{d{\bar\omega}(r)}{dr} \right] =
16\pi G A^4(\varphi)(\rho + p){\bar\omega}(r).
\end{eqnarray}

The natural boundary conditions for ${\bar\omega}$ are

\begin{eqnarray}
\frac{d{\bar\omega}}{dr}(0)= 0 \;\;\; {\rm and}\;\;\; \lim_{r\to \infty}{\bar\omega}=\Omega .
\end{eqnarray}
The first condition ensures the regularity of $\bar\omega$ at the center of the star.

One of the quantities we consider in the present paper is the inertial moment $I$ of the  compact star. It is defined as usual

\begin{eqnarray}
I=\frac{J}{\Omega}.
\end{eqnarray}
Using  equation (\ref{OR}) for $\bar \omega$  and the asymptotic form of $\bar \omega$ one can also show that

\begin{eqnarray}\label{eq:I_integral}
I= \frac{8\pi G}{3} \int_{0}^{r_S}A^4(\varphi)(\rho + p)e^{\Lambda - \Phi} r^4 \left(\frac{\bar\omega}{\Omega}\right) dr .
\end{eqnarray}

In the next section where we present our numerical results we shall use the dimensionless parameter $m_{\varphi}\to m_{\varphi} R_{0}$  and the dimensionless inertial moment
$I\to I/M_{\odot}R^2_{0} $ where $M_{\odot}$ is the solar mass and $R_{0}=1.47664 \,km$ is one half of the solar gravitational radius.

\section{Results}
\subsection{Constraints on the parameters of the theory}\label{Sec:Constraints}
Let us first discuss the case of massless scalar field with coupling function $A(\varphi) = \exp{(\frac{1}{2}\beta\varphi^2)}$ which is equivalent to general relativity in the weak field regime and therefore, passes without any problem through most of the observations. The only exception are the observation of neutron stars in close binaries where the strong field effects are non-negligible. These binary systems consist normally of two neutron stars or a neutron star and  white dwarf and the most complete list of such objects up to our knowledge can be found in \cite{Freire2012} including the recently discovered PSR J0348+0432 \cite{Antoniadis13}. Since the emitted gravitational waves match very well the predictions of general relativity, constraints on $\beta$ can be obtained from the requirement to have negligible amount of scalar gravitational radiation for the corresponding binary system (this means that we should have very weakly scalarized or completely non-scalarized solutions for the observed neutron star masses) \cite{Damour1996}. All of the observed binary systems lead to constraints on $\beta>-5.0$ and the most severe bound comes from PSR J0348+0432, namely $\beta > -4.5$. This narrows down significantly the possible range of $\beta$ since scalarization is observed roughly for $\beta<-4.35$ in the static and $\beta<-3.9$ in the rapidly rotating case \cite{Doneva2013}.

A way to circumvent this severe constraint is to consider a massive scalar field as discussed in the introduction \cite{Popchev2015,Ramazanouglu2016}. The mass of the scalar field can effectively suppress the scalar gravitational waves and reconcile the scalar-tensor theories with the binary neutron star observations for a much larger range of $\beta$. In more rigorous terms, if the Compton wave-length of the scalar field $\lambda_\varphi$ is much smaller than the separation of the two stars in the binary system denoted with $r_b$, the emitted scalar gravitational radiation will be negligible. Therefore, the strongest bound on the $\lambda_\varphi$, and thus on the scalar field mass $m_\varphi$, would come from the binary system in \cite{Freire2012,Antoniadis13} which have the smallest orbital separation. It turns out though that the orbital separation for all these systems is roughly of the same order $10^9 {\rm m}$. This translated into
\begin{equation}
m_\varphi \gg 10^{-16} {\rm eV}.
\end{equation}
As far as neutron stars are concerned for such values of $m_{\varphi}$, $\beta$ is practically unconstrained. Additional bounds on $\beta$ were given in \cite{Ramazanouglu2016}, namely $-10^3 < \beta < -3 $. In this section we will present results for $\beta \ge -10$, since much smaller values of $\beta$ lead to very drastic changes in the neutron star structure.

An upper limit on $m_\varphi$ can be imposed based on the requirement that the mass term does not prevent the scalarization of the star. Namely, the characteristic length scale of the star should be smaller than the Compton wave-length which leads to $m_\varphi\lesssim 10^{-9}{\rm eV}$ \cite{Ramazanouglu2016}. Therefore the allowed range for $m_\varphi$ is
\begin{equation} \label{eq:bounds_mphi}
10^{-16} {\rm eV} \lesssim m_\varphi \lesssim 10^{-9}{\rm eV}.
\end{equation}
A mid-range can be also excluded ($10^{-13} {\rm eV} \lesssim m_\varphi \lesssim 10^{-11}{\rm eV}$) based on the arguments connected to superradiant instability if we assume that the measurement of the black hole spin is accurate enough \cite{Cardoso2013a,Ramazanouglu2016}. There are many uncertainties in this case though and that is why we will consider the whole range of $m_\varphi$ given by eq. \eqref{eq:bounds_mphi} in order to achieve a completeness of our results.

Let us now turn to the bounds imposed on the parameters in the massive Brans-Dicke case. In the massless case the coupling parameter $\alpha_0$ in eq. \eqref{eq:BD_couplingFunction} is severely limited since the Brans-Dicke theory gives deviations from GR even in the weak field regime that is tested with high accuracy by several experiments. If the scalar field is massive enough though it will be exponentially suppressed and would not influence these experiment. As we commented above, the tightest constraint on the scalar field mass would come from the experiment where we have the smallest separation. From the macrophysics test, this is the Gravity Probe B experiments, where the separation is equal to the orbit of the satellite that is of the order of $10^7 {\rm m}$. Other observations, such as the advance of Mercury perihelion or the deflection of light by the Sun, have characteristic length scale of the order of one astronomical unit, that is clearly much larger.

Thus, if we impose the requirement that the Compton wave-length is smaller than $10^7 {\rm m}$, we obtains the following constraint for the massive Brans-Dicke theory
\begin{equation}
m_{\varphi} > 2 \times 10^{-14} eV.
\end{equation}
Using such values of the scalar mass, the parameter $\alpha_0$ is essentially unconstrained.

\subsection{Numerical results}
The system of reduced field equations is solved using a shooting method, where the shooting parameters are the values of the scalar field and the metric functions $\Phi$ and $\omega$ at the origin. Once we fix the mass of the scalar field and the coupling parameters $\beta$ or $\alpha_0$ (depending on the particular class of STT we are using), the solution is specified by the central energy density $\rho_c$. The code is tested against the results in the static case presented in \cite{Ramazanouglu2016} and it shows very good agreement.

In the case when $A(\varphi) = \exp{(\frac{1}{2}\beta\varphi^2)}$ we have non-uniqueness of the solutions. The trivial case with zero scalar field (i.e. the GR case) is always a solution of the field equations but for certain regions of the parameter space there are additional solutions with the same central energy density but different nonzero scalar fields. In the case of massive Brans-Dicke theory with $A(\varphi) = \exp(\alpha_0 \varphi)$ we have unique solutions specified by the central energy density and the value of $\alpha_0$. The trivial case with zero scalar field is not a solution of the field equations for nonzero $\alpha_0$.

We use a representative modern equation of state, the so-called APR EOS and the piecewise polytropic approximation is employed \cite{Read2013}.

\subsubsection{Massive scalar-tensor theory with $A(\varphi) = \exp(\frac{1}{2}\beta\varphi^2)$ which admits scalarization.}
\begin{figure}[]
	\centering
	\includegraphics[width=0.9\textwidth]{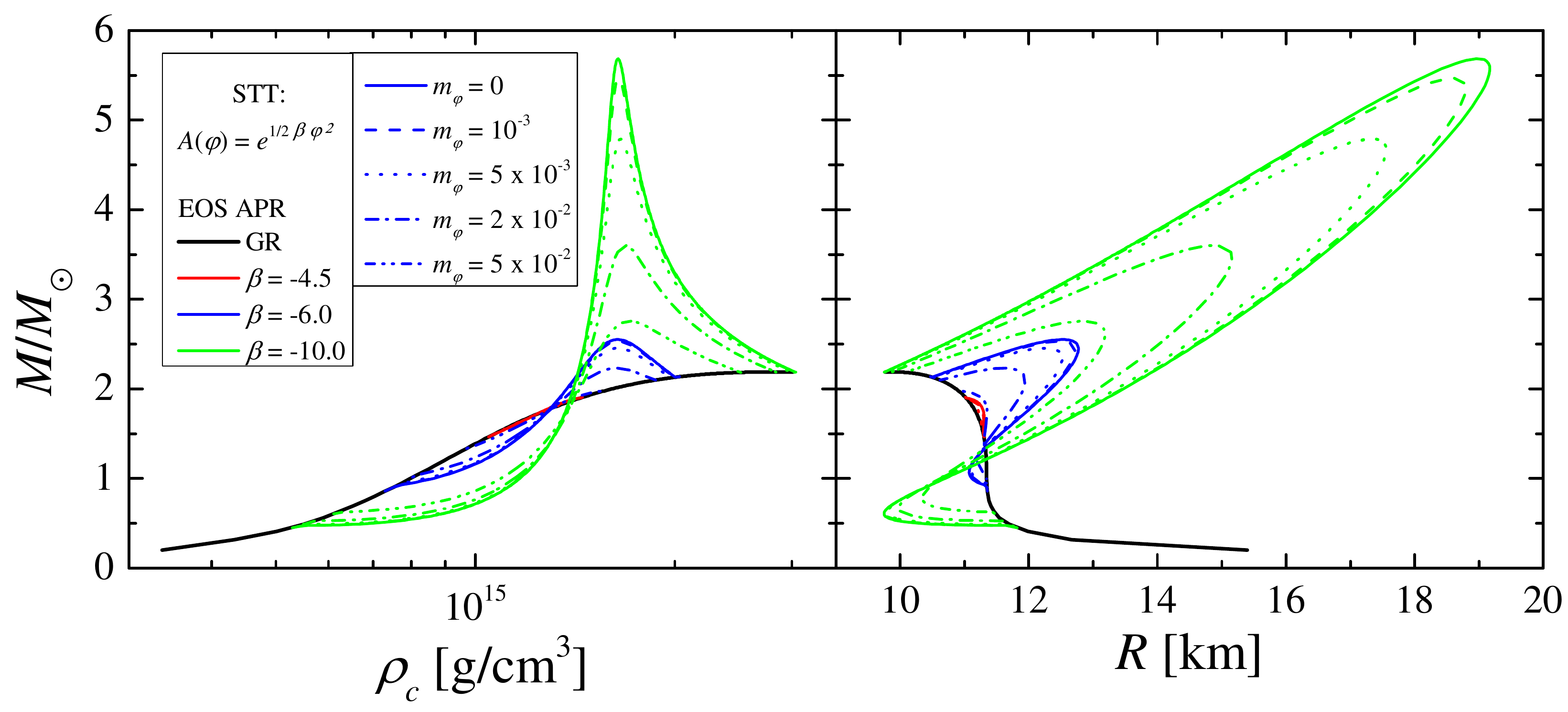}
	\caption{The mass as a function of the central energy density (left panel) and as a function of the radius (right panel) for EOS APR. The results for different values of the coupling constant $\beta$ and mass of the scalar field $m_\varphi$ are plotted.}
	\label{Fig:MR}
\end{figure}

As we have already commented, in this case the observations of close binary systems can not impose constraints on the parameter $\beta$ if the mass of the scalar field is sufficiently large $m_\varphi \gg 10^{-16} {\rm eV}$. In Fig. \ref{Fig:MR} the mass as a function of the central energy density and the radius is presented for several combinations of $\beta$ and the scalar field mass $m_\varphi$. The presented results in this figure are for the static case since we are calculating only first order corrections with respect to the stellar rotational frequency $\Omega$ while the rotational corrections to the mass $M$ and the radius $R$ are of second order of $\Omega$. In the figure we plot the data for $\beta \ge -10$ and one can see that for such values of the coupling parameter the maximum neutron star mass increases almost 3 times and the stellar radii are also reach significantly larger values. If we increase $\beta$ even more we will naturally get larger deviations from GR, but the qualitative behavior will remain the same.

Expectedly, the results for $\beta=-4.5$ differ only marginally from  the GR case, but the differences can reach very large values with the increase of $\beta$. The case with $m_\varphi=0$ is clearly equivalent to the massless STT. As we increase the scalar field mass the Compton wave-length decreases. This effectively suppresses the scalar field and the deviation from GR start to decrease. Loosely speaking in the limit when $m_\varphi \rightarrow \infty$ the solutions converge to the GR ones. Therefore, the neutron stars with different values of $m_\varphi$ are more or less bounded between the massless STT case and the GR limit. This means that the scalar field mass can not lead to larger deviation from the pure GR compared to the massless theory for the same value of $\beta$ but instead what we gain is the much broader range of allowed values of $\beta$. In the figure we have plotted the results for masses between zero and $5 \times 10^{-2}$ in our dimensionless units. Our calculations show that if we increase the $m_\varphi$ further the solutions get closer and closer to the pure GR, but we decided not to plot them, because they will overlap with the solutions for different values of $\beta$, thus making the figure overcrowded and difficult to read. For masses smaller than $10^{-3}$ the results become almost indistinguishable from the massless case.

\begin{figure}[]
	\centering
	\includegraphics[width=0.45\textwidth]{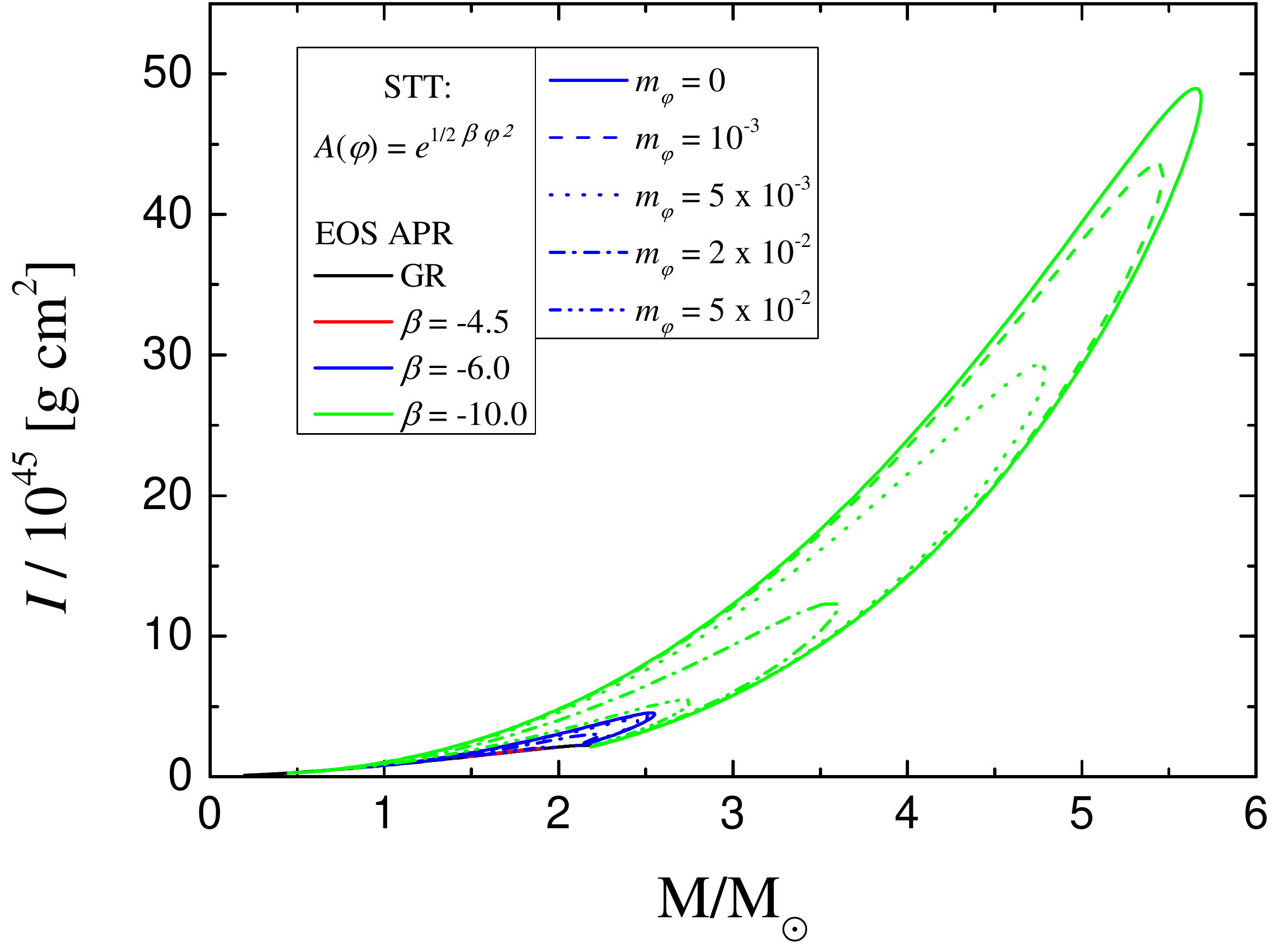}
	\includegraphics[width=0.45\textwidth]{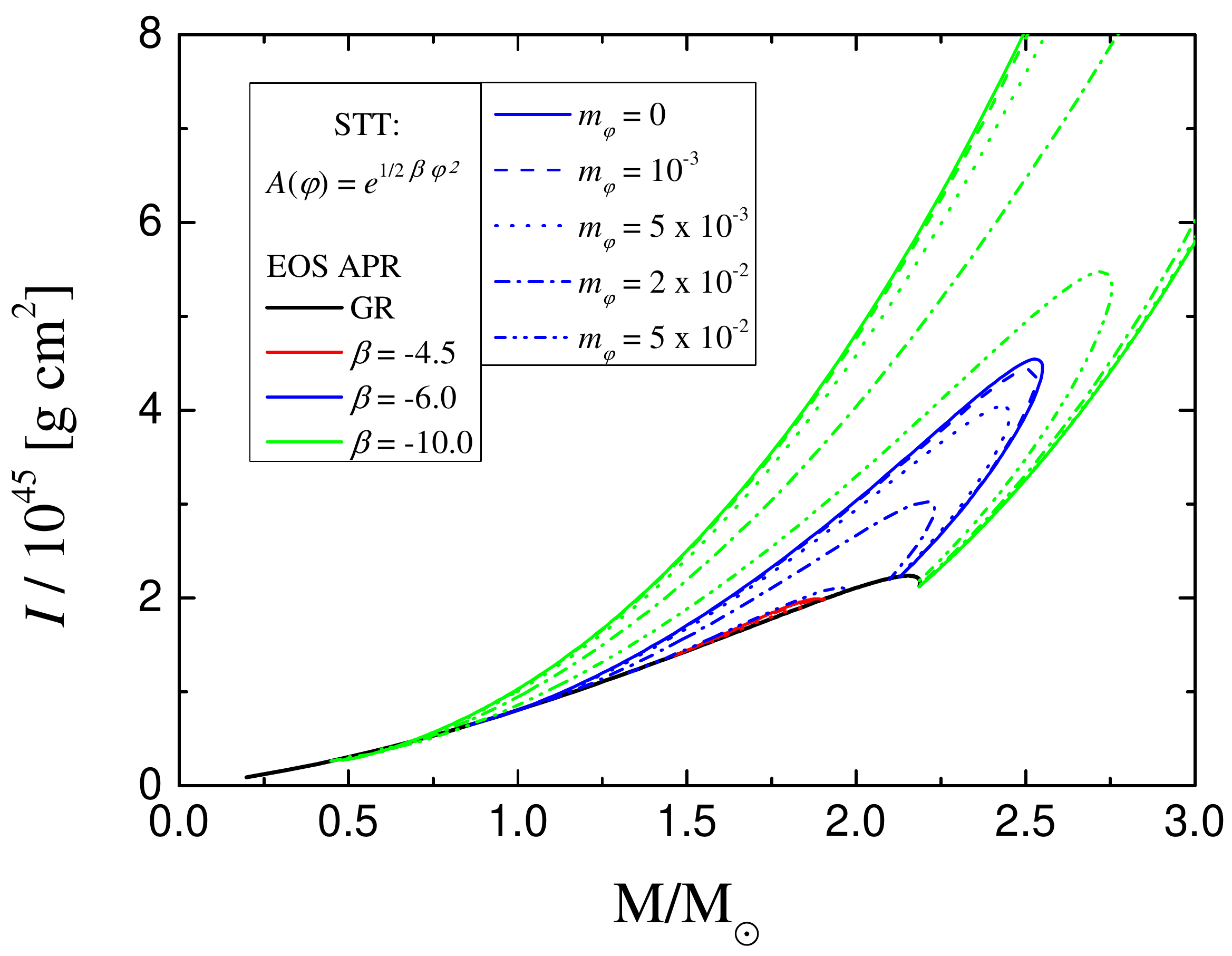}
	\caption{The moment of inertia as a function of the stellar mass. The right panel is a magnification of the left panel.}
	\label{Fig:IM}
\end{figure}

In Fig. \ref{Fig:IM} the moment of inertia $I$ is plotted as a function of the stellar mass, where the right panel is a magnification of the left one. As one can see $I$ can increase by almost an order of magnitude compared to the general relativistic case for $\beta=10$. This is really a drastic change that can be used to impose observational constraint on the massive STT since it is expected that in the near future the moment of inertia of binary neutron stars will be observed with a good accuracy \cite{Lattimer2005}. Again we should note that if we increase further the scalar field mass, the results will get close to the GR case this overlapping with the results for larger $\beta$. This is the reason why we decided not to plot data for larger values of $m_\varphi$

Let us comment further on the change of the results when varying $\beta$ and $m_\varphi$. As one can see the qualitative behavior of both the mass and the moment of inertia is practically the same and the results with small $\beta$ and large $m_\varphi$ practically overlap with the data for large $\beta$ and small $m_\varphi$. This means that observations of the stellar mass, radius and moment of inertia alone can give us a clue for possible deviations from GR but can not discriminate between effects coming from varying $\beta$ and $m_\varphi$. Such discrimination might be possible with other observations such as the gravitational wave signal emitted by inspiraling binaries \cite{Barausse2013,Palenzuela2014,Shibata2014}. On the other hand if close binary systems are discovered that have a significant smaller orbital separation compared to the ones already observed \cite{Freire2012,Antoniadis13}, then the mass of the scalar field can be further constrained as discussed in Section \ref{Sec:Constraints}.

Here we consider only one equation of state that can be viewed as a representative one. A common problem for testing the alternative theories of gravity is that there is a degeneracy between effects coming from varying the gravitational theory and the equation of state. As one can see here the situation is much better because the deviations from GR can be dramatic while the parameters are still in agreement with all the observations.

The values of $m_\varphi$ presented on the graph are in agreement with the bounds discussed in Section \ref{Sec:Constraints} coming from the observations of close binary systems and the requirement that the mass term does not prevent the scalarization of the star, namely $10^{-16} {\rm eV} \lesssim m_\varphi \lesssim 10^{-9}{\rm eV}$ (in our dimensionless units $7\times 10^{-7} \lesssim m_\varphi  \lesssim 7 $). We have not excluded though in our calculations the mid-range constraint coming from measurements of black hole spins and the superradiant instability $10^{-13} {\rm eV} \lesssim m_\varphi \lesssim 10^{-11}{\rm eV}$ (in our dimensionless units $7\times 10^{-4} \lesssim m_\varphi  \lesssim 7 \times 10^{-2}$). The reason is that we wanted to achieve a completeness of our studies and of course there are uncertainties in the measurements of the black holes spins that can lead to a change of this mid-range.

\subsubsection{Massive Brans-Dicke theory with $A(\varphi) = \exp(\alpha_0 \varphi)$.}
\begin{figure}[]
	\centering
	\includegraphics[width=0.9\textwidth]{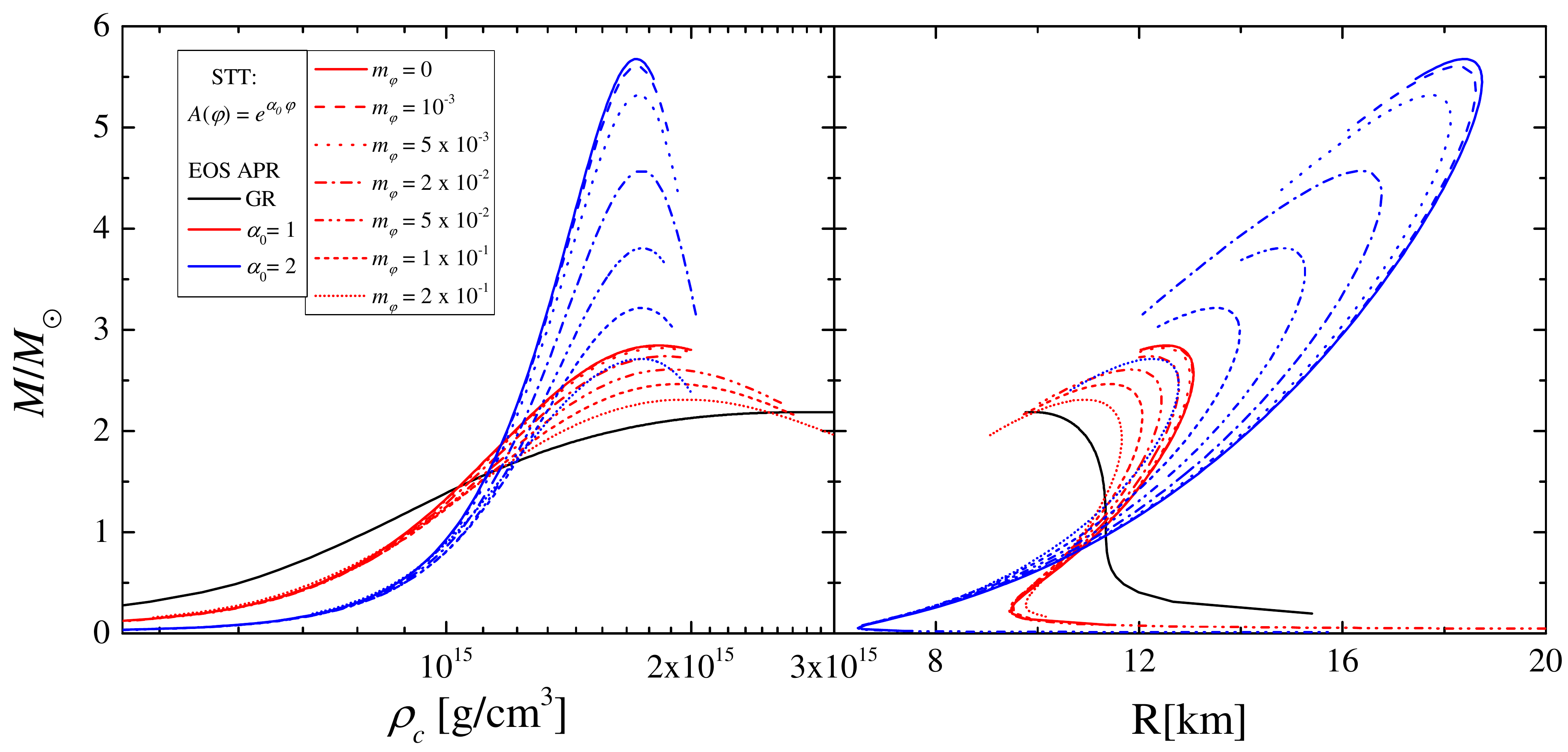}
	\caption{The mass as a function of the central energy density (left panel) and as a function of the radius (right panel) for neutron stars in the massive Brans-Dicke theory with EOS APR. The results for different values of the coupling constant $\alpha_0$ and mass of the scalar field $m_\varphi$ are plotted.}
	\label{Fig:MR_BD}
\end{figure}

\begin{figure}[]
	\centering
	\includegraphics[width=0.6\textwidth]{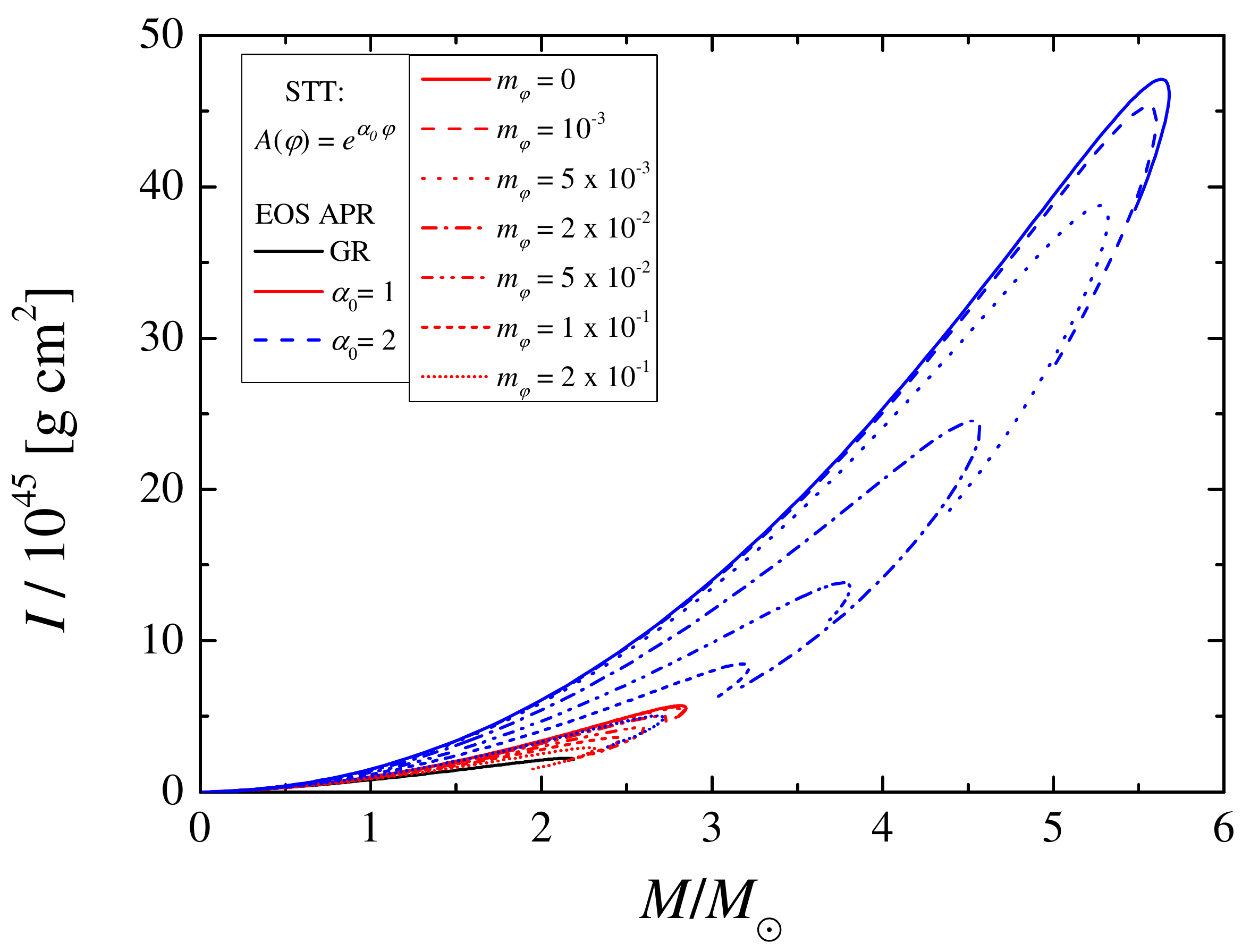}
	\caption{The moment of inertia as a function of the neutron star mass for the massive Brans-Dicke theory}
	\label{Fig:IM_BD}
\end{figure}

As a second class of scalar-tensor theory we will consider the massive Brans-Dicke theory. As we discussed earlier, the tightest bound on the scalar field mass comes from the Gravity Probe B experiment, namely $m_\varphi > 10^{-4}$. The results are presented in Figs. \ref{Fig:MR_BD} and \ref{Fig:IM_BD}. We have chosen two values of $\alpha_0$, $\alpha_0=1$ and $\alpha_0=2$ which lead to a significant change of the neutron star properties. As one can see for nonzero $\alpha_0$ the solutions always differ from GR, unlike in the previous section, which is a straightforward consequence from the field equations and the exact form of the coupling function. For the considered values of $\alpha_0$ the maximum mass can increases almost three times and the moment of inertia increases by roughly an order of magnitude compared to the GR case. With the increase of $\alpha_0$ the deviations from the Einstein's theory of gravity get even larger.

As one can see the qualitative behavior is similar to the STT with coupling function $A(\varphi) = \exp(\frac{1}{2}\beta\varphi^2)$ considered in the previous section. For a fixed value of $\alpha_0$, the solutions with different scalar field masses are bounded between the general relativistic solutions, which correspond to the limit of infinite scalar field mass, and the massless Brans-Dicke solutions. The presented results are for $m_\varphi \ge 10^{-4}$ in agreement with the observations and it can be noticed that for $m_\varphi = 10^{-4}$ the results are already very close to the massless Brans-Dicke case. Another similarity with the results in the previous section is that the qualitative behavior of changing the parameter $\alpha_0$ and the scalar field mass is the same. This means that solutions with small $\alpha_0$ and small $m_\varphi$ overlap with the solutions with large $\alpha_0$ and large $m_\varphi$ which makes it very difficult to distinguish between the two effects.

We will not go into further details in the results for this particular class of STT for the following reasons. First, the results are qualitatively very similar to the results in the previous section as we commented and most of the conclusions made there are valid also for the massive Brans-Dicke theory. Second, our main goal here was not to make an extensive study but rather to give representative examples that the massive Brans-Dicke theory can lead to significant deviations for GR for values of the coupling parameters that are in agreement with present observations, contrary to the massless Brans-Dicke theory where the deviations from GR are marginal because of the tight constraints imposed by the weak-field observations.

\section{Conclusions}
We have studied slowly rotating neutron stars in two particular classes of scalar-tensor theory with nonzero scalar field mass. The first one is equivalent to GR in the weak field regime for massless scalar field, but it can lead to large deviations when strong field are considered (the so-called scalarization). The second one is the massive Brans-Dicke theory. These two classes are amongst the most intuitive and widely used STTs.

In both theories the static and slowly neutron star solutions differ almost marginally from GR in the massless case if one considers coupling parameters that are in agreement with the present observation. The inclusion of scalar field mass changes the picture dramatically. It suppresses the scalar field at length scale of the order of the Compton wavelength which helps us reconcile the theory with the observations for a much broader range of the coupling parameters. Indeed, it turns out that the mass, radius and moment of inertia for neutron stars in massive STT can differ drastically from the pure GR solutions if sufficiently large masses of the scalar field are considered.

One inconvenience comes from the fact that the effects of changing the coupling parameters and the scalar field mass have the same qualitative influence on several important dependences such as mass-radius and moment of inertia-mass, that are considered in the present paper. Therefore, one can not break the degeneracy between these parameters using only observations of the neutron star mass, radius and moment of inertia. For this purpose different astrophysical implications have to be considered, such as the emitted gravitational wave signal after binary neutron star merger, similar to the massless case \cite{Barausse2013,Palenzuela2014,Shibata2014}.

The considered massive STTs are ones of the very few (if not the only) currently available alternative theories of gravity that have very large deviations from GR on one hand and be in agreement with all the present observations on the other. In addition, the scalar-tensor theory is a very well posed theory which does not suffer from intrinsic problems and it is one of the most natural generalizations of Einstein's theory of gravity. This makes the results in the present paper and their astrophysical implications important and worth exploring further in order to propose specific tests for constraining the parameters of the theory.

\section*{Acknowledgements}
SY would like to thank the Research Group Linkage Programme of the Alexander von Humboldt Foundation for the support. DD would like to thank the European Social Fund and the Ministry of Science, Research and the Arts Baden-W\"urttemberg for the support. The support by the Bulgarian NSF Grant DFNI T02/6 and "New-CompStar" COST Action MP1304 is also  gratefully acknowledged.


\bibliography{references}

\begin{thebibliography}{23}
\expandafter\ifx\csname natexlab\endcsname\relax\def\natexlab#1{#1}\fi
\expandafter\ifx\csname bibnamefont\endcsname\relax
  \def\bibnamefont#1{#1}\fi
\expandafter\ifx\csname bibfnamefont\endcsname\relax
  \def\bibfnamefont#1{#1}\fi
\expandafter\ifx\csname citenamefont\endcsname\relax
  \def\citenamefont#1{#1}\fi
\expandafter\ifx\csname url\endcsname\relax
  \def\url#1{\texttt{#1}}\fi
\expandafter\ifx\csname urlprefix\endcsname\relax\def\urlprefix{URL }\fi
\providecommand{\bibinfo}[2]{#2}
\providecommand{\eprint}[2][]{\url{#2}}

\bibitem[{\citenamefont{{Damour} and {Esposito-Farese}}(1993)}]{Damour1993}
\bibinfo{author}{\bibfnamefont{T.}~\bibnamefont{{Damour}}} \bibnamefont{and}
  \bibinfo{author}{\bibfnamefont{G.}~\bibnamefont{{Esposito-Farese}}},
  \bibinfo{journal}{Physical Review Letters} \textbf{\bibinfo{volume}{70}},
  \bibinfo{pages}{2220} (\bibinfo{year}{1993}).

\bibitem[{\citenamefont{{Damour} and {Esposito-Far{\`e}se}}(1996)}]{Damour1996}
\bibinfo{author}{\bibfnamefont{T.}~\bibnamefont{{Damour}}} \bibnamefont{and}
  \bibinfo{author}{\bibfnamefont{G.}~\bibnamefont{{Esposito-Far{\`e}se}}},
  \bibinfo{journal}{\prd} \textbf{\bibinfo{volume}{54}}, \bibinfo{pages}{1474}
  (\bibinfo{year}{1996}).

\bibitem[{\citenamefont{{Harada}}(1997)}]{Harada1997}
\bibinfo{author}{\bibfnamefont{T.}~\bibnamefont{{Harada}}},
  \bibinfo{journal}{Progress of Theoretical Physics}
  \textbf{\bibinfo{volume}{98}}, \bibinfo{pages}{359} (\bibinfo{year}{1997}).

\bibitem[{\citenamefont{{Harada}}(1998)}]{Harada1998}
\bibinfo{author}{\bibfnamefont{T.}~\bibnamefont{{Harada}}},
  \bibinfo{journal}{\prd} \textbf{\bibinfo{volume}{57}}, \bibinfo{pages}{4802}
  (\bibinfo{year}{1998}).

\bibitem[{\citenamefont{{Salgado} et~al.}(1998)\citenamefont{{Salgado},
  {Sudarsky}, and {Nucamendi}}}]{Salgado1998}
\bibinfo{author}{\bibfnamefont{M.}~\bibnamefont{{Salgado}}},
  \bibinfo{author}{\bibfnamefont{D.}~\bibnamefont{{Sudarsky}}},
  \bibnamefont{and}
  \bibinfo{author}{\bibfnamefont{U.}~\bibnamefont{{Nucamendi}}},
  \bibinfo{journal}{\prd} \textbf{\bibinfo{volume}{58}}, \bibinfo{eid}{124003}
  (\bibinfo{year}{1998}).

\bibitem[{\citenamefont{{Pani} et~al.}(2011)\citenamefont{{Pani}, {Macedo},
  {Crispino}, and {Cardoso}}}]{Pani2011}
\bibinfo{author}{\bibfnamefont{P.}~\bibnamefont{{Pani}}},
  \bibinfo{author}{\bibfnamefont{C.~F.~B.} \bibnamefont{{Macedo}}},
  \bibinfo{author}{\bibfnamefont{L.~C.~B.} \bibnamefont{{Crispino}}},
  \bibnamefont{and}
  \bibinfo{author}{\bibfnamefont{V.}~\bibnamefont{{Cardoso}}},
  \bibinfo{journal}{\prd} \textbf{\bibinfo{volume}{84}}, \bibinfo{eid}{087501}
  (\bibinfo{year}{2011}).

\bibitem[{\citenamefont{{Sotani}}(2012)}]{Sotani2012}
\bibinfo{author}{\bibfnamefont{H.}~\bibnamefont{{Sotani}}},
  \bibinfo{journal}{\prd} \textbf{\bibinfo{volume}{86}}, \bibinfo{eid}{124036}
  (\bibinfo{year}{2012}).

\bibitem[{\citenamefont{{Doneva} et~al.}(2013)\citenamefont{{Doneva},
  {Yazadjiev}, {Stergioulas}, and {Kokkotas}}}]{Doneva2013}
\bibinfo{author}{\bibfnamefont{D.~D.} \bibnamefont{{Doneva}}},
  \bibinfo{author}{\bibfnamefont{S.~S.} \bibnamefont{{Yazadjiev}}},
  \bibinfo{author}{\bibfnamefont{N.}~\bibnamefont{{Stergioulas}}},
  \bibnamefont{and} \bibinfo{author}{\bibfnamefont{K.~D.}
  \bibnamefont{{Kokkotas}}}, \bibinfo{journal}{\prd}
  \textbf{\bibinfo{volume}{88}}, \bibinfo{eid}{084060} (\bibinfo{year}{2013}).

\bibitem[{\citenamefont{{Freire} et~al.}(2012)\citenamefont{{Freire}, {Wex},
  {Esposito-Far{\`e}se}, {Verbiest}, {Bailes}, {Jacoby}, {Kramer}, {Stairs},
  {Antoniadis}, and {Janssen}}}]{Freire2012}
\bibinfo{author}{\bibfnamefont{P.~C.~C.} \bibnamefont{{Freire}}},
  \bibinfo{author}{\bibfnamefont{N.}~\bibnamefont{{Wex}}},
  \bibinfo{author}{\bibfnamefont{G.}~\bibnamefont{{Esposito-Far{\`e}se}}},
  \bibinfo{author}{\bibfnamefont{J.~P.~W.} \bibnamefont{{Verbiest}}},
  \bibinfo{author}{\bibfnamefont{M.}~\bibnamefont{{Bailes}}},
  \bibinfo{author}{\bibfnamefont{B.~A.} \bibnamefont{{Jacoby}}},
  \bibinfo{author}{\bibfnamefont{M.}~\bibnamefont{{Kramer}}},
  \bibinfo{author}{\bibfnamefont{I.~H.} \bibnamefont{{Stairs}}},
  \bibinfo{author}{\bibfnamefont{J.}~\bibnamefont{{Antoniadis}}},
  \bibnamefont{and} \bibinfo{author}{\bibfnamefont{G.~H.}
  \bibnamefont{{Janssen}}}, \bibinfo{journal}{\mnras}
  \textbf{\bibinfo{volume}{423}}, \bibinfo{pages}{3328} (\bibinfo{year}{2012}).

\bibitem[{\citenamefont{Antoniadis et~al.}(2013)\citenamefont{Antoniadis,
  Freire, Wex, Tauris, Lynch et~al.}}]{Antoniadis13}
\bibinfo{author}{\bibfnamefont{J.}~\bibnamefont{Antoniadis}},
  \bibinfo{author}{\bibfnamefont{P.~C.} \bibnamefont{Freire}},
  \bibinfo{author}{\bibfnamefont{N.}~\bibnamefont{Wex}},
  \bibinfo{author}{\bibfnamefont{T.~M.} \bibnamefont{Tauris}},
  \bibinfo{author}{\bibfnamefont{R.~S.} \bibnamefont{Lynch}},
  \bibnamefont{et~al.}, \bibinfo{journal}{Science}
  \textbf{\bibinfo{volume}{340}}, \bibinfo{pages}{6131} (\bibinfo{year}{2013}).

\bibitem[{\citenamefont{{Perivolaropoulos}}(2010)}]{Perivolaropoulos2010}
\bibinfo{author}{\bibfnamefont{L.}~\bibnamefont{{Perivolaropoulos}}},
  \bibinfo{journal}{\prd} \textbf{\bibinfo{volume}{81}}, \bibinfo{eid}{047501}
  (\bibinfo{year}{2010}), \eprint{0911.3401}.

\bibitem[{\citenamefont{{Alsing} et~al.}(2012)\citenamefont{{Alsing}, {Berti},
  {Will}, and {Zaglauer}}}]{Alsing2012}
\bibinfo{author}{\bibfnamefont{J.}~\bibnamefont{{Alsing}}},
  \bibinfo{author}{\bibfnamefont{E.}~\bibnamefont{{Berti}}},
  \bibinfo{author}{\bibfnamefont{C.~M.} \bibnamefont{{Will}}},
  \bibnamefont{and}
  \bibinfo{author}{\bibfnamefont{H.}~\bibnamefont{{Zaglauer}}},
  \bibinfo{journal}{\prd} \textbf{\bibinfo{volume}{85}}, \bibinfo{eid}{064041}
  (\bibinfo{year}{2012}), \eprint{1112.4903}.

\bibitem[{\citenamefont{{Ramazano{\u g}lu} and
  {Pretorius}}(2016)}]{Ramazanouglu2016}
\bibinfo{author}{\bibfnamefont{F.~M.} \bibnamefont{{Ramazano{\u g}lu}}}
  \bibnamefont{and}
  \bibinfo{author}{\bibfnamefont{F.}~\bibnamefont{{Pretorius}}},
  \bibinfo{journal}{ArXiv e-prints}  (\bibinfo{year}{2016}),
  \eprint{1601.07475}.

\bibitem[{\citenamefont{Popchev}(July 2015)}]{Popchev2015}
\bibinfo{author}{\bibfnamefont{D.}~\bibnamefont{Popchev}}, Master's thesis,
  \bibinfo{school}{University of Sofia} (\bibinfo{year}{July 2015}).

\bibitem[{\citenamefont{{Chen} et~al.}(2015)\citenamefont{{Chen}, {Suyama}, and
  {Yokoyama}}}]{Chen2015}
\bibinfo{author}{\bibfnamefont{P.}~\bibnamefont{{Chen}}},
  \bibinfo{author}{\bibfnamefont{T.}~\bibnamefont{{Suyama}}}, \bibnamefont{and}
  \bibinfo{author}{\bibfnamefont{J.}~\bibnamefont{{Yokoyama}}},
  \bibinfo{journal}{\prd} \textbf{\bibinfo{volume}{92}}, \bibinfo{eid}{124016}
  (\bibinfo{year}{2015}).

\bibitem[{\citenamefont{{Hartle}}(1967)}]{Hartle1967}
\bibinfo{author}{\bibfnamefont{J.~B.} \bibnamefont{{Hartle}}},
  \bibinfo{journal}{\apj} \textbf{\bibinfo{volume}{150}}, \bibinfo{pages}{1005}
  (\bibinfo{year}{1967}).

\bibitem[{\citenamefont{Yazadjiev et~al.}(2014)\citenamefont{Yazadjiev, Doneva,
  Kokkotas, and Staykov}}]{Yazadjiev2014}
\bibinfo{author}{\bibfnamefont{S.~S.} \bibnamefont{Yazadjiev}},
  \bibinfo{author}{\bibfnamefont{D.~D.} \bibnamefont{Doneva}},
  \bibinfo{author}{\bibfnamefont{K.~D.} \bibnamefont{Kokkotas}},
  \bibnamefont{and} \bibinfo{author}{\bibfnamefont{K.~V.}
  \bibnamefont{Staykov}}, \bibinfo{journal}{JCAP}
  \textbf{\bibinfo{volume}{1406}}, \bibinfo{pages}{003} (\bibinfo{year}{2014}).

\bibitem[{\citenamefont{{Cardoso} et~al.}(2013)\citenamefont{{Cardoso},
  {Carucci}, {Pani}, and {Sotiriou}}}]{Cardoso2013a}
\bibinfo{author}{\bibfnamefont{V.}~\bibnamefont{{Cardoso}}},
  \bibinfo{author}{\bibfnamefont{I.~P.} \bibnamefont{{Carucci}}},
  \bibinfo{author}{\bibfnamefont{P.}~\bibnamefont{{Pani}}}, \bibnamefont{and}
  \bibinfo{author}{\bibfnamefont{T.~P.} \bibnamefont{{Sotiriou}}},
  \bibinfo{journal}{Physical Review Letters} \textbf{\bibinfo{volume}{111}},
  \bibinfo{eid}{111101} (\bibinfo{year}{2013}).

\bibitem[{\citenamefont{{Read} et~al.}(2013)\citenamefont{{Read}, {Baiotti},
  {Creighton}, {Friedman}, {Giacomazzo}, {Kyutoku}, {Markakis}, {Rezzolla},
  {Shibata}, and {Taniguchi}}}]{Read2013}
\bibinfo{author}{\bibfnamefont{J.~S.} \bibnamefont{{Read}}},
  \bibinfo{author}{\bibfnamefont{L.}~\bibnamefont{{Baiotti}}},
  \bibinfo{author}{\bibfnamefont{J.~D.~E.} \bibnamefont{{Creighton}}},
  \bibinfo{author}{\bibfnamefont{J.~L.} \bibnamefont{{Friedman}}},
  \bibinfo{author}{\bibfnamefont{B.}~\bibnamefont{{Giacomazzo}}},
  \bibinfo{author}{\bibfnamefont{K.}~\bibnamefont{{Kyutoku}}},
  \bibinfo{author}{\bibfnamefont{C.}~\bibnamefont{{Markakis}}},
  \bibinfo{author}{\bibfnamefont{L.}~\bibnamefont{{Rezzolla}}},
  \bibinfo{author}{\bibfnamefont{M.}~\bibnamefont{{Shibata}}},
  \bibnamefont{and}
  \bibinfo{author}{\bibfnamefont{K.}~\bibnamefont{{Taniguchi}}},
  \bibinfo{journal}{ArXiv e-prints}  (\bibinfo{year}{2013}).

\bibitem[{\citenamefont{{Lattimer} and {Schutz}}(2005)}]{Lattimer2005}
\bibinfo{author}{\bibfnamefont{J.~M.} \bibnamefont{{Lattimer}}}
  \bibnamefont{and} \bibinfo{author}{\bibfnamefont{B.~F.}
  \bibnamefont{{Schutz}}}, \bibinfo{journal}{\apj}
  \textbf{\bibinfo{volume}{629}}, \bibinfo{pages}{979} (\bibinfo{year}{2005}).

\bibitem[{\citenamefont{{Barausse} et~al.}(2013)\citenamefont{{Barausse},
  {Palenzuela}, {Ponce}, and {Lehner}}}]{Barausse2013}
\bibinfo{author}{\bibfnamefont{E.}~\bibnamefont{{Barausse}}},
  \bibinfo{author}{\bibfnamefont{C.}~\bibnamefont{{Palenzuela}}},
  \bibinfo{author}{\bibfnamefont{M.}~\bibnamefont{{Ponce}}}, \bibnamefont{and}
  \bibinfo{author}{\bibfnamefont{L.}~\bibnamefont{{Lehner}}},
  \bibinfo{journal}{\prd} \textbf{\bibinfo{volume}{87}}, \bibinfo{eid}{081506}
  (\bibinfo{year}{2013}).

\bibitem[{\citenamefont{{Palenzuela} et~al.}(2014)\citenamefont{{Palenzuela},
  {Barausse}, {Ponce}, and {Lehner}}}]{Palenzuela2014}
\bibinfo{author}{\bibfnamefont{C.}~\bibnamefont{{Palenzuela}}},
  \bibinfo{author}{\bibfnamefont{E.}~\bibnamefont{{Barausse}}},
  \bibinfo{author}{\bibfnamefont{M.}~\bibnamefont{{Ponce}}}, \bibnamefont{and}
  \bibinfo{author}{\bibfnamefont{L.}~\bibnamefont{{Lehner}}},
  \bibinfo{journal}{\prd} \textbf{\bibinfo{volume}{89}}, \bibinfo{eid}{044024}
  (\bibinfo{year}{2014}).

\bibitem[{\citenamefont{{Shibata} et~al.}(2014)\citenamefont{{Shibata},
  {Taniguchi}, {Okawa}, and {Buonanno}}}]{Shibata2014}
\bibinfo{author}{\bibfnamefont{M.}~\bibnamefont{{Shibata}}},
  \bibinfo{author}{\bibfnamefont{K.}~\bibnamefont{{Taniguchi}}},
  \bibinfo{author}{\bibfnamefont{H.}~\bibnamefont{{Okawa}}}, \bibnamefont{and}
  \bibinfo{author}{\bibfnamefont{A.}~\bibnamefont{{Buonanno}}},
  \bibinfo{journal}{\prd} \textbf{\bibinfo{volume}{89}}, \bibinfo{eid}{084005}
  (\bibinfo{year}{2014}).

\end{thebibliography}

\end{document}